\documentclass[prb,reprint,aps,showpacs,showkeys,twocolumn,superscriptaddress]{revtex4-1}
\pdfoutput=1

\usepackage[utf8]{inputenc}
\usepackage[T1]{fontenc}
\usepackage{hyperref}
\usepackage{color}
\usepackage{graphicx}
\usepackage{units}
\usepackage{amsmath}
\usepackage{dcolumn}
\newcolumntype{d}[1]{D{.}{.}{#1}}

\hypersetup{colorlinks = true, citecolor = blue, breaklinks = true}

\begin{document}

\title{Strain-induced structural instability in FeRh}
\date{\today}
\author{Ulrich Aschauer}
\affiliation{Materials Theory, ETH Zurich, Wolfgang-Pauli-Strasse 27, CH-8093 Z\"urich, Switzerland}
\affiliation{Department of Chemistry and Biochemistry, University of Bern, Freiestrasse 3, CH-3012 Bern, Switzerland}
\author{Roisin Braddell}
\affiliation{Materials Theory, ETH Zurich, Wolfgang-Pauli-Strasse 27, CH-8093 Z\"urich, Switzerland}
\author{Sonia A. Brechb\"uhl}
\affiliation{Materials Theory, ETH Zurich, Wolfgang-Pauli-Strasse 27, CH-8093 Z\"urich, Switzerland}
\author{Peter M. Derlet}
\affiliation{Condensed Matter Theory Group, Paul Scherrer Institute, CH-5232 Villigen PSI, Switzerland}
\author{Nicola A. Spaldin}
\affiliation{Materials Theory, ETH Zurich, Wolfgang-Pauli-Strasse 27, CH-8093 Z\"urich, Switzerland}

\begin{abstract}
We perform density functional calculations to investigate the structure of the inter-metallic alloy FeRh under epitaxial strain. Bulk FeRh exhibits a metamagnetic transition from a low-temperature antiferromagnetic (AFM) phase to a ferromagnetic (FM) phase at 350K, and its strain dependence is of interest for tuning the transition temperature to the room-temperature operating conditions of typical memory devices. We find an unusually strong dependence of the structural energetics on the choice of exchange-correlation functional, with the usual local density approximation (LDA) yielding the wrong ground-state structure, and generalized gradient (GGA) extensions being in better agreement with the bulk experimental structure. Using the GGA we show the existence of a metastable face-centered-cubic (fcc)-like AFM structure that is reached from the ground state body-centered-cubic (bcc) AFM structure by following the epitaxial Bain path. We predict that this metastable fcc-like structure has a significantly higher conductivity than the bcc AFM phase. We show that the behavior is well described using non-linear elasticity theory, which captures the softening and eventual sign change of the orthorhombic shear modulus under compressive strain, consistent with this structural instability. Finally, we predict the existence of an additional unit-cell-doubling lattice instability, which should be observable at low temperature. 
\end{abstract}

\maketitle

\section{Introduction}

\begin{figure}
\includegraphics[width=0.9\columnwidth]{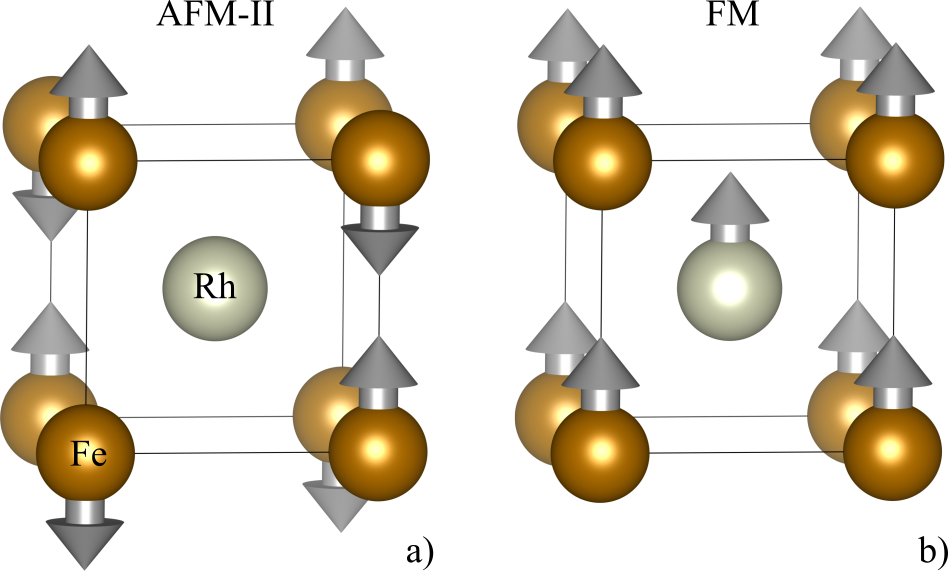}
\caption{\label{fig:struct} Magnetic phases of FeRh: a) antiferromagnetic (AFM-II) and b) ferromagnetic order.}
\end{figure}

The inter-metallic compound FeRh exhibits an unusual first-order phase transition from an antiferromagnetic (AFM) structure at low temperature to a ferromagnetic (FM) structure above roughly 350 K \cite{Fallot:1939uk,Kouvel:1962ba}. The phase transition is isostructural, with both phase having the body-centered cubic (bcc) CsCl structure. The AFM state (Fig. \ref{fig:struct}a) is characterised by magnetic moments at the Fe sites of around 3$\mu_B$ and negligible moments at the Rh sites, while the FM state (Fig. \ref{fig:struct}b) shows slightly larger moments (around 3.3$\mu_B$) on the Fe sites and Rh also acquires a moment of about 1$\mu_B$ \cite{Shirane:1964tl}. The phase transition is accompanied by an increase in volume of about 1\% and a large drop in resistivity \cite{Kouvel:1962ba}.

In recent years interest in FeRh has been rejuvenated following its growth in thin-film form, and the associated potential for integration into device architectures. A number of technologically promising behaviors exploiting the coupling between magnetism, resistivity and volume / strain have been demonstrated. These include magnetic-field writing with resistive reading in FeRh films grown on MgO \cite{Marti:2014fl}, small voltage switching between AFM and FM states for FeRh on ferroelectric BaTiO$_3$ \cite{Cherifi:2014du} and electric-field phase and resistivity control for FeRh on piezoelectric PMN-PT \cite{Lee:2015jj}.

A number of density functional theory (DFT) calculations have already been performed to investigate the properties of FeRh. Early work based on the local density approximation (LDA) studied the relative energetics of different magnetic phases and found the ground state to be of type AFM-II (we will refer to this as AFM in the following) \cite{Moruzzi:1992vz}. At larger than equilibrium volumes the FM state is energetically preferred within the LDA, which is consistent with the volume increase associated with the AFM to FM phase transition. Subsequent work argued that in analogy to Fe, gradient corrected functionals (GGA) are required for the correct description of FeRh \cite{Gruner:2003kl}. It was shown that while LDA and GGA functionals predict similar structures, the magnetic moments on Fe in the AFM phase are significantly larger and in better agreement with experiment using GGA \cite{Lounis:2003hn}. Later work showed that close to the transition volume, the Fe moments are unstable with respect to canting when treated at the LDA level whereas a semi-local GGA functional stabilises the collinear FM phase \cite{Gu:2005jk}. Using non-collinear LDA calculations it was argued that the Rh moment cannot be described within the Stoner picture as the spin density shows spatial variations around the Rh atom \cite{Sandratskii:2011by}. This study also established the importance of strong Fe-Rh hybridisation and showed that the Fe-Fe antiferromagnetic interaction is strongly volume dependent, whereas the ferromagnetic Fe-Rh interaction remains fairly constant with volume.

Following the recent interest in coherent epitaxial strained FeRh films, some density functional studies have been carried out for strained systems. In terms of electrical properties, DFT calculations suggested that injecting holes does not significantly affect the relative stability of the AFM and the FM phases, whereas injecting about 0.4 electrons per formula unit makes the FM phase energetically favourable compared to the AFM phase \cite{Cherifi:2014du}. This was confirmed by calculations of the actual interface between ferroelectric BaTiO$_3$ and FeRh, in which polarised BaTiO$_3$ injects electrons and leads to a favouring of the FM state. The FM magnetic ordering was predicted to have a significantly higher density of states at the Fermi energy, and hence a higher electrical conductivity, than the AFM ordering \cite{Szajek:1994cx,Gu:2005jk,Sandratskii:2011by,Cherifi:2014du,Lee:2015jj}. No significant change of the density of states at the Fermi energy was found, however, for small ($\pm 0.1$ \%) strain ranges \cite{Lee:2015jj}. It was also shown that for very thin films (9 layers) the FM state is energetically favoured over the AFM state \cite{Lounis:2003hn}. Intriguingly, and relevant for the work that we present here,  Ref.~\onlinecite{Cherifi:2014du} showed an unusual energy lowering for the AFM phase under compressive strain at the LDA level, although they did not comment on its existence or its origin. 

Here we report results of detailed DFT calculations of the {\it structure} of FeRh under strain. We are motivated in part by a lesser known experimental fact about FeRh, namely that FeRh converts to a face-centered cubic (fcc) structure when subjected to high velocity impact deformation or other strong deformation such as filing \cite{Lommel:1967cc,Miyajima:1992ta}. The fcc structure reverts to the body-centered cubic (bcc) FM CsCl structure on heating above 500K and finally to the AFM structure when cooling to room temperature, suggesting a Bain path transformation between the bcc and fcc structures. Indeed, it has been shown using LDA TB-LMTO calculations that the AFM fcc state has a lower energy than the FM bcc state \cite{Pugacheva:1994wp}, but a complete investigation of the Bain path in the low temperature AFM phase is lacking.

Our main finding is that epitaxial strain can convert FeRh from the ground-state AFM bcc phase to a metastable AFM fcc-like phase by following the epitaxial Bain path. The behavior manifests in the non-linear elastic constants as a softening of the orthorhombic shear distortion in the AFM phase under for compressive strain. We show that the detailed energetics of the conversion is strongly dependent on the choice of exchange-correlation functional. Using the GGA, which yields better agreement with experiment, the transition from bcc to fcc-like structure has a very small energy barrier and the two phases have very similar energies. We predict that the conductivities of the two phases will be very different, based on their different densities of states at the Fermi energy. Finally, our calculations reveal a phonon instability for the AFM bcc structure, which we predict will lead to a dimerisation of both Fe and Rh at low temperature. 

\section{Computational details}

Our density functional theory calculations were performed using the VASP code \cite{Kresse:1993ty,Kresse:1994us,Kresse:1996vk,Kresse:1996vf} with both the LDA and PBE \cite{Perdew:1996iq} density functionals. The influence of correlations was studied using the DFT+U method \cite{Anisimov:1991wt} by applying a rotationally invariant U on the Fe d states \cite{Dudarev:1998vn}. Wave functions were expanded in plane waves up to a kinetic energy cutoff of 550 eV. PAW potentials \cite{Blochl:1994uk,Kresse:1999wc} with Fe(4s, 3d, 3p) and Rh(5s, 4d, 4p) treated as valence electrons were used. Reciprocal space was sampled using a 19x19x19 gamma-centred mesh for the 2 atom unit cell and a 11x11x11 gamma-centred mesh for the 2x2x2 16 atom supercell. These values yield well-converged results for forces, stress and phonon frequencies. Phonon calculations were performed using the frozen phonon method in the phonopy code \cite{Togo:2008jt}. Biaxial strain was applied by constraining the length of two of the axes and varying the length of the third axis to find its lowest energy value.

\section{Results \& Discussion}
\subsection{Dependence of properties on the exchange-correlation functional}

Before proceeding to calculate the strain-dependent structure of FeRh, we make an extensive comparison of the properties obtained for the bulk system using different exchange-correlation functionals, both with literature data from other DFT calculations, as well as with experiment. 

\subsubsection{Structure and magetism}

\begin{figure}
\includegraphics[width=0.9\columnwidth]{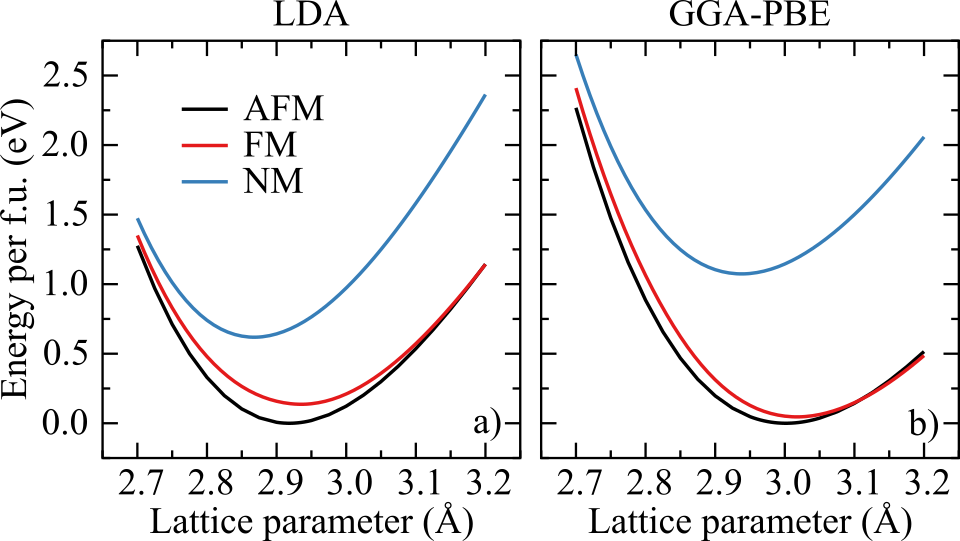}
\caption{\label{fig:isostatic}Energy vs. cubic lattice parameter for the non-magnetic (NM, blue), antiferromagnetic (AFM, black) and ferromagnetic (FM, red) phases of FeRh.}
\end{figure}

In figure \ref{fig:isostatic} we show the energy vs. cubic lattice parameter for the non-magnetic (NM), antiferromagnetic (AFM) and ferromagnetic (FM) phases in the cubic CsCl structure computed using the LDA (panel a) and GGA (panel b). (Note that the NM case is obtained by not allowing spin-polarization in the calculation.) Table \ref{tab:minima} compares our optimised lattice parameters and relative energies with results of previously published calculations.

\begin{table*}
\caption{Lattice constants, energies per formula unit relative to the AFM phase and magnetic moments computed in this work and from the literature.}
\begin{ruledtabular}
\begin{tabular}{lld{1.3}d{3.0}cd{1.3}d{3.0}c}
Phase	& & \multicolumn{3}{c}{LDA} & \multicolumn{3}{c}{GGA} \\\cline{3-5}\cline{6-8}
                & & \multicolumn{1}{c}{Lattice} & \multicolumn{1}{c}{Energy} & \multicolumn{1}{c}{Magn. mom.} & \multicolumn{1}{c}{Lattice} & \multicolumn{1}{c}{Energy} & \multicolumn{1}{c}{Magn. mom.}\\
		& & \multicolumn{1}{c}{(\AA)} & \multicolumn{1}{c}{(meV)} & \multicolumn{1}{c}{Fe, Rh ($\mu_B$)} & \multicolumn{1}{c}{(\AA)} & \multicolumn{1}{c}{(meV)} & \multicolumn{1}{c}{Fe, Rh ($\mu_B$)}\\
\hline
AFM		& This work					& 2.918	& -		& 2.85, 0.00			& 3.002	& -		& 3.15, 0.00			\\
                & Ref. \onlinecite {Moruzzi:1992vz}	& 2.989	& -		& 2.98, 0.00			&		&		&					\\
		& Ref. \onlinecite{Szajek:1994cx}	& 3.009	& -		& 3.13, 0.00			& 		&		&					\\
                & Ref. \onlinecite{Gruner:2003kl}   	&		&		&					& 3.002	& -		& 3.18, 0.00			\\
                & Ref. \onlinecite{Lounis:2003hn}  	&		&		&					& 2.974	& -		& 3.11, 0.00			\\
                & Ref. \onlinecite{Gu:2005jk}		& 2.973	& -		& 3.12, 0.00			& 2.996	& -		& 3.28, 0.00			\\
		& Ref. \onlinecite{Cherifi:2014du}	& 2.922	& -		& 3.21, 0.00			&		&		&					\\
\hline
FM		& This work					& 2.935	& 136	& 2.98, 1.00			& 3.018	& 46		& 3.21, 1.05			\\
                & Ref. \onlinecite {Moruzzi:1992vz}	& 3.006	& 52		& 3.15, 1.02			&		&		&					\\
		& Ref. \onlinecite{Szajek:1994cx}	& 3.020	& 60		& 3.20, 1.02			&		&		&					\\
                & Ref. \onlinecite{Gruner:2003kl}   	&		&		&					& 3.020	& 68		& 3.23, 1.00			\\
                & Ref. \onlinecite{Lounis:2003hn}  	&		&		&					& 2.990	&		& 3.19, 1.05			\\
                & Ref. \onlinecite{Gu:2005jk}		& 2.987	& 51		& 3.22, 1.04			& 3.018	& 6		& 3.31, 1.02			\\
		& Ref. \onlinecite{Cherifi:2014du}	& 2.935	& 93		& 3.29, 0.94			&		&		&					\\
\hline
NM		& This work					& 2.868	& 619	&					& 2.938	& 1074	& 					\\
                & Ref. \onlinecite {Moruzzi:1992vz}	&		&		&					&		&		&					\\
		& Ref. \onlinecite{Szajek:1994cx}	& 2.958	& 756	&					&		&		&					\\
                & Ref. \onlinecite{Gruner:2003kl}   	&		&		&					& 2.959	& 1088	& 					\\
\end{tabular}
\end{ruledtabular}
\label{tab:minima}
\end{table*}

According to experiment, the FM phase has a larger lattice parameter (2.999 \AA) than the AFM phase (2.984 \AA) \cite{Shirane:1963vb,Shirane:1964tl}. We see this also realised in our results - independent of the density functional - as well as all previous computational studies. Also independent of the functional, the AFM phase is predicted to have a lower energy than the FM phase, a fact that was also reported by all previous DFT studies and is in agreement with the experimental low temperature AFM ground state. In general the values obtained from our calculations agree with previous reports, the agreement being better for GGA, while our LDA calculations predict around 6\% smaller lattice parameters than previous calculations. We notice, however, that very similar lattice parameters to ours were found by the most recent LDA study \cite{Cherifi:2014du}.

In table \ref{tab:minima} we also report the predicted magnetic moments. Our computed moments agree well with previously reported values at the LDA\cite{Moruzzi:1992vz,Gu:2005jk} and GGA \cite{Gruner:2003kl,Lounis:2003hn,Gu:2005jk} level, where values of around 3 $\mu_B$ for the Fe and 1 $\mu_B$ for the Rh moments where reported. These data show that the computed LDA local Fe magnetic moments are about 0.2 to 0.3 $\mu_B$ smaller than the GGA values for both the AFM and FM phase. 

We make three observations at this point: 1) There is a spread of -6.6\% to +2.5\% with respect to experiment in the LDA lattice parameters and from -1.0\% to +2.1\% in the GGA lattice parameters respectively. A general rule of thumb is that LDA systematically underestimates lattice parameters, while GGA overestimates them. This rule is not observed here, hinting at an extreme sensitivity of the obtained lattice parameters to the density functional and other details of the DFT calculations. 2) The FM phase is predicted to be at least 46 meV per formula unit higher than the AFM phase, except for the GGA calculations of Ref.~\onlinecite{Gu:2005jk} where an energy difference of only 6 meV was found. Experimentally this energy difference has been determined to be 5.34 meV per formula unit \cite{Ponomarev:1973uv}, which shows that in general DFT calculations tend to overestimate this energy difference. 3) Under compression, both functionals tend towards a cross-over to a non-magnetic state, indicating the presence of a magneto-volume effect.

\subsubsection{Linear elastic constants} \label{ss:ec}

To further asses the performance of different density functionals, we report in table \ref{tab:constants} the elastic constants computed by fitting to isostatic as well as volume-conserving orthorhombic and tetragonal strain in a range of -5 to +5\% \cite{Mehl:1994wu}. We show values for the FM and AFM phases, as well as NM for completeness. For the AFM phase at the LDA level, the range had to be reduced to -2.5\% to +2.5\% in order to obtain a good fit in the orthorhombic strain case, hinting already that these second-order elastic constants do not fully capture the elastic behavior of FeRh and that we might anticipate unusual strain-dependent behavior. 

Experimentally, the AFM phase (ground state in a $\mathrm{Fe_{0.98}Rh_{1.02}}$ alloy) was measured to have a slightly higher bulk modulus of 141 GPa than the FM phase (ground state in $\mathrm{Fe_{1.04}Rh_{0.96}}$) with 133 GPa \cite{Cooke:2012jl}. Despite the slightly different composition, we expect the elastic constants in $\mathrm{Fe_{1.00}Rh_{1.00}}$ to be comparable. Our predicted bulk moduli are higher than these experimental values, especially at the LDA level, where the values are overestimated by a factor of 1.60 and 1.74 respectively for the AFM and FM phases. The GGA predictions are closer to experiment but still overestimate by a factor 1.36 and 1.45 respectively. Despite this overestimation, it is encouraging that the AFM phase is predicted to have a higher bulk modulus than the FM phase. Our results are also consistent with previous predictions at the LDA level (AFM: 214.4 GPa \cite{Moruzzi:1992vz}, 226.8 GPa \cite{Szajek:1994cx}, 245.4 GPa \cite{Gu:2005jk}; FM: 201.6 GPa \cite{Moruzzi:1992vz}, 244.0 GPa \cite{Szajek:1994cx}, 236.4 GPa \cite{Gu:2005jk}) as well as the GGA level (AFM: 197 GPa \cite{Gruner:2003kl}, 219.4 GPa \cite{Gu:2005jk}; FM: 193 GPa \cite{Gruner:2003kl}, 218.1 GPa \cite{Gu:2005jk}).

\begin{table}
\caption{Elastic constants (in GPa) for the different magnetic phases of FeRh computed using the LDA and GGA-PBE functionals.}
\begin{ruledtabular}
\begin{tabular}{lld{4.3}d{4.3}}
\multicolumn{2}{l}{Phase} & \multicolumn{1}{c}{LDA} & \multicolumn{1}{c}{GGA-PBE} \\
\hline
AFM		& $B$				& 246.802	& 195.978 \\
		& $C_{11}-C_{12}$		& -12.850 	& 44.206 \\
		& $C_{11}$ 			& 238.236 	& 225.449 \\
		& $C_{12}$			& 251.086 	& 181.243 \\ 
		& $C_{44}$			& 132.891 	& 119.867 \\
\hline
FM		& $B$ 				& 213.441	& 193.298 \\
		& $C_{11}-C_{12}$		& 66.626		& 87.397 \\
		& $C_{11}$ 			& 257.858	& 251.563 \\
		& $C_{12}$ 			& 191.232	& 164.166 \\ 
		& $C_{44}$ 			& 118.751	& 109.773 \\
\hline
NM		& $B$				& 295.075	& 239.574 \\
		& $C_{11}-C_{12}$		& -314.345	& -262.455 \\
		& $C_{11}$ 			& 85.512		& 64.604 \\
		& $C_{12}$ 			& 399.857 	& 327.059 \\ 
		& $C_{44}$ 			& 166.448	& 137.492 \\
\end{tabular}
\end{ruledtabular}
\label{tab:constants}
\end{table}

The data in table \ref{tab:constants} have two particularly interesting implications, given that a material is elastically stable only when $C_{11}-C_{12}$ > 0. First, we see that, independent of the functional, the NM phase does not satisfy this criterion and is hence elastically unstable, implying that FeRh is in fact structurally stabilized by its magnetic interactions. It is somewhat surprising, however, that the AFM phase is also elastically unstable when using the LDA functional whereas it is stable using GGA. This incomplete stabilization could be attributed to the smaller magnetic moments, which yield reduced magnetic interactions  at the LDA level. The FM phase on the other hand is elastically stable independent of the functional. 

We conclude at this point that based on structural parameters, elastic properties and size of the magnetic moments, the GGA yields a better description of bulk FeRh than the LDA. In what follows, therefore we use PBE-GGA exclusively for calculating the strain-dependent structure. Since we focus on the ground-state AFM ordering, the overestimate of the energy difference to the FM phase, which is one inherent problem of the PBE functional, will not affect our conclusions, but we emphasize that care should be taken when considering the relative stability of the different magnetic orderings using DFT calculations.

\subsection{Epitaxial strain dependence} \label{ss:esd}

\begin{figure}
\includegraphics[width=0.9\columnwidth]{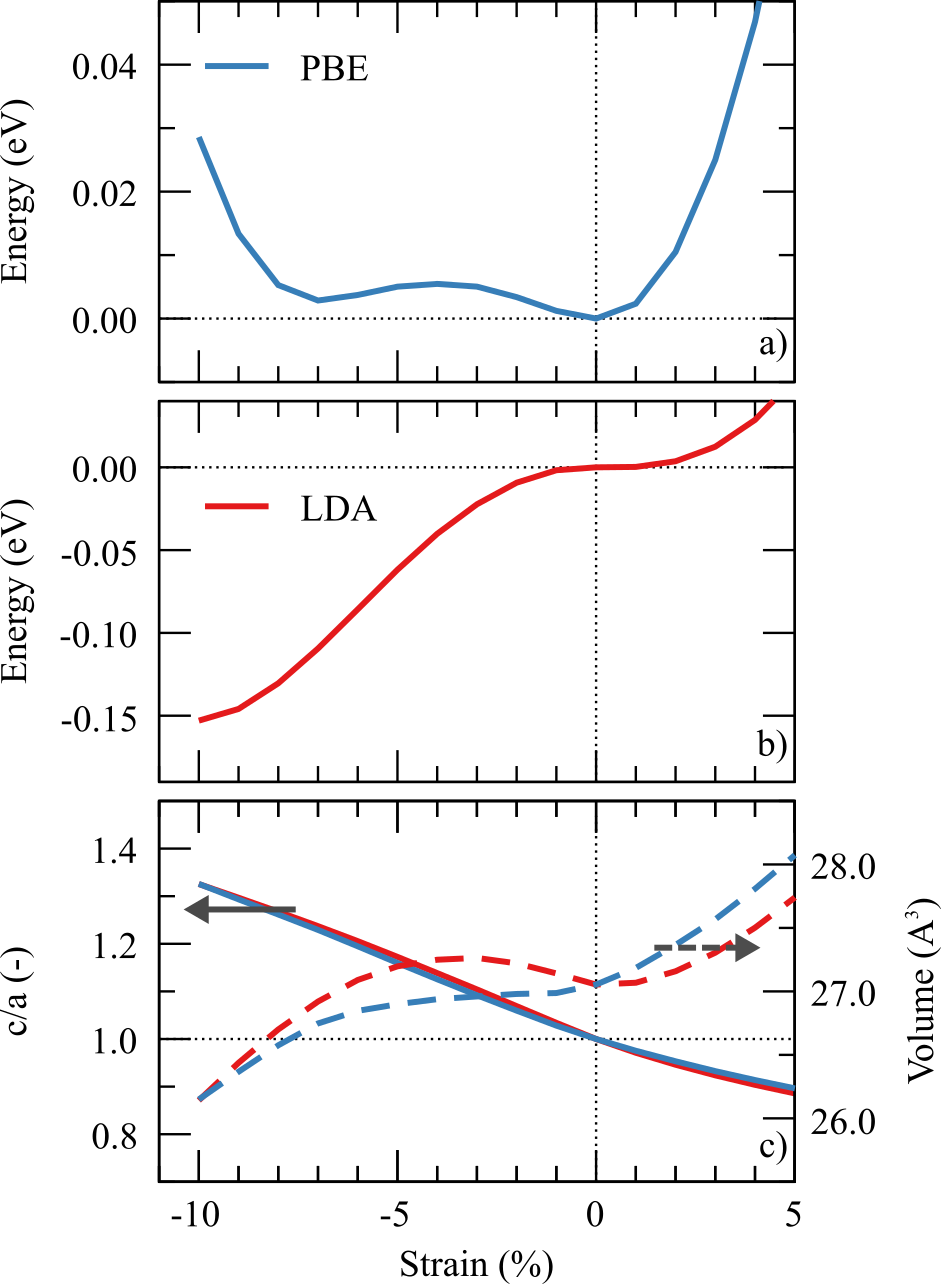}
\caption{\label{fig:strain}Calculated strain dependence of the energy of the AFM phase computed at a) the GGA-PBE and b) the LDA level along with c) their respective c/a ratios (solid lines, left axis) and volume (dashed lines, right axis).}
\end{figure}

In figure \ref{fig:strain} a) we show the energy evolution of the AFM phase as a function of epitaxial strain for the PBE functional and for sake of completeness also for the LDA functional. While under tensile strain both functionals predict the expected approximately quadratic increase in energy with increasing strain, the functionals behave markedly differently under compressive strain.

Using LDA (Fig. \ref{fig:strain}b), the bcc CsCl structure of the AFM phase is merely an inflection point in the energy landscape and we observe a significant energy lowering under compressive epitaxial strain, which is a manifestation of the negative $C_{11}-C_{12}$ value reported in table \ref{tab:constants}. A similar energy evolution was previously shown but not commented upon \cite{Cherifi:2014du}. Accompanying the reduction in energy is an increase in the c/a ratio (Fig. \ref{fig:strain}b) towards $\sqrt{2}=1.414$, indicating that a transition along the tetragonal Bain path has taken FeRh from the bcc CsCl structure to an fcc-like structure. 

At the GGA level (Fig. \ref{fig:strain}a), which we concluded above yields a better description of FeRh, the energy of the AFM phase increases under compressive strain and the bcc CsCl structure is predicted to be the ground state. There exists however a metastable state at -7\% strain, only 3 meV/f.u. higher in energy than the bcc CsCl structure and reached from the bcc phase by overcoming a small energy barrier of 5 meV/f.u. This is in agreement with the experimental finding that plastic deformation can induce a transformation from bcc to fcc, which is reversed upon heating \cite{Lommel:1967cc,Miyajima:1992ta}.

Another implication of this result is that  FeRh should exhibit a very non-standard behaviour under compressive strain. At 0\% strain and around 7\% strain, one should consider the material unstrained, but existing in two distinct phases (bcc and fcc-like respectively). If grown at intermediate strains, the material could relax into either of the two phases as the film grows thicker, which could lead to phase coexistence in partially relaxed films, reminiscent of the different phases occurring in BiFeO$_3$ \cite{Zeches_et_al:2009}. Fig. \ref{fig:strain}c) reveals that the volume change is minimal for the strain range encompassing the two energy minima, further supporting a phase coexistence scenario. 

\begin{figure}
\includegraphics[width=0.9\columnwidth]{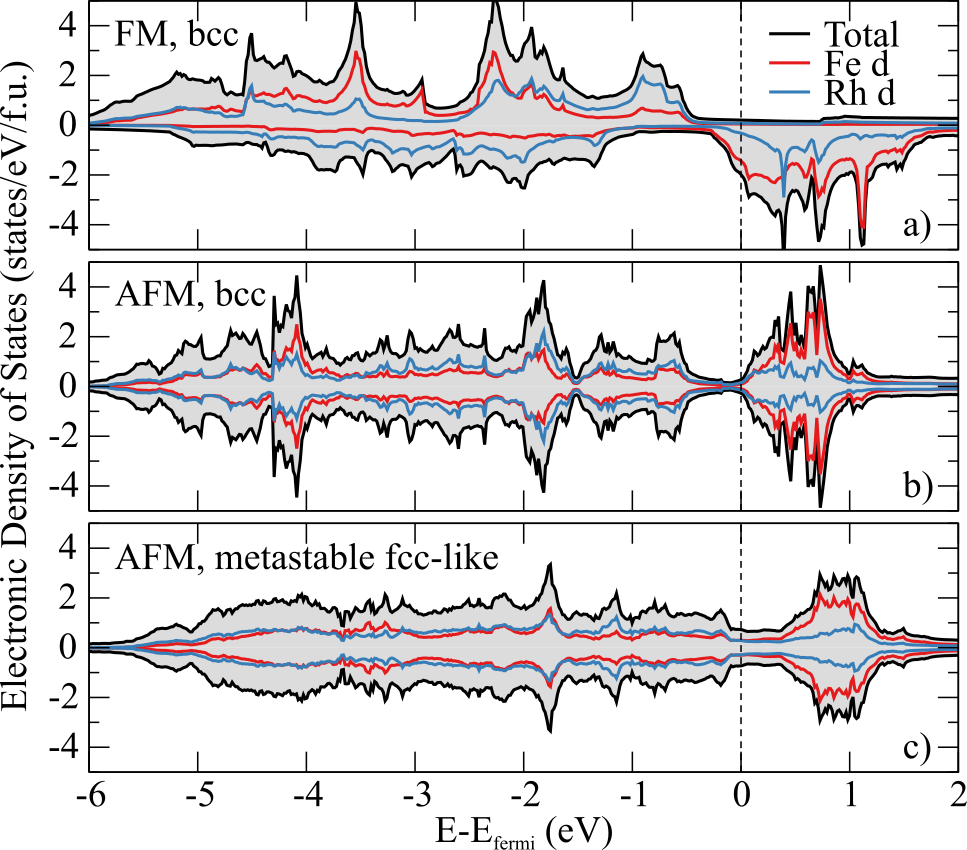}
\caption{\label{fig:dos}Electronic densities of states (DOS) for a) the bcc FM, b) the bcc AFM and c) the metastable fcc-like AFM phases.}
\end{figure}

In Fig. \ref{fig:dos} we show the GGA computed electronic densities of states (DOSs) for the bcc a) FM and b) AFM phases respectively as well as c) the metastable fcc-like AFM structure at  7\% compressive strain. All DOSs are reported per formula unit of FeRh so that the DOS at the Fermi energy can be directly compared for the different structures. The much larger DOS at the Fermi energy for the bcc FM phase compared to the bcc AFM phase is in agreement with previous calculations \cite{Gruner:2003kl,Lounis:2003hn,Gu:2005jk}. Comparing the DOS for the metastable fcc-like AFM structure (Fig. \ref{fig:dos}c) it is interesting to note that this structure has almost as large a DOS at the Fermi energy as the FM phase, when considering both up and down spin electrons. An enhancement of the DOS at the Fermi energy was previously predicted to occur in the FM phase under compressive strain \cite{Pugacheva:1994wp}. Here we predict that a change from the AFM bcc to the AFM fcc-like structure will result in a change in resistivity comparable to that accompanying the AFM-FM phase transition.

\begin{figure}
\includegraphics[width=0.9\columnwidth]{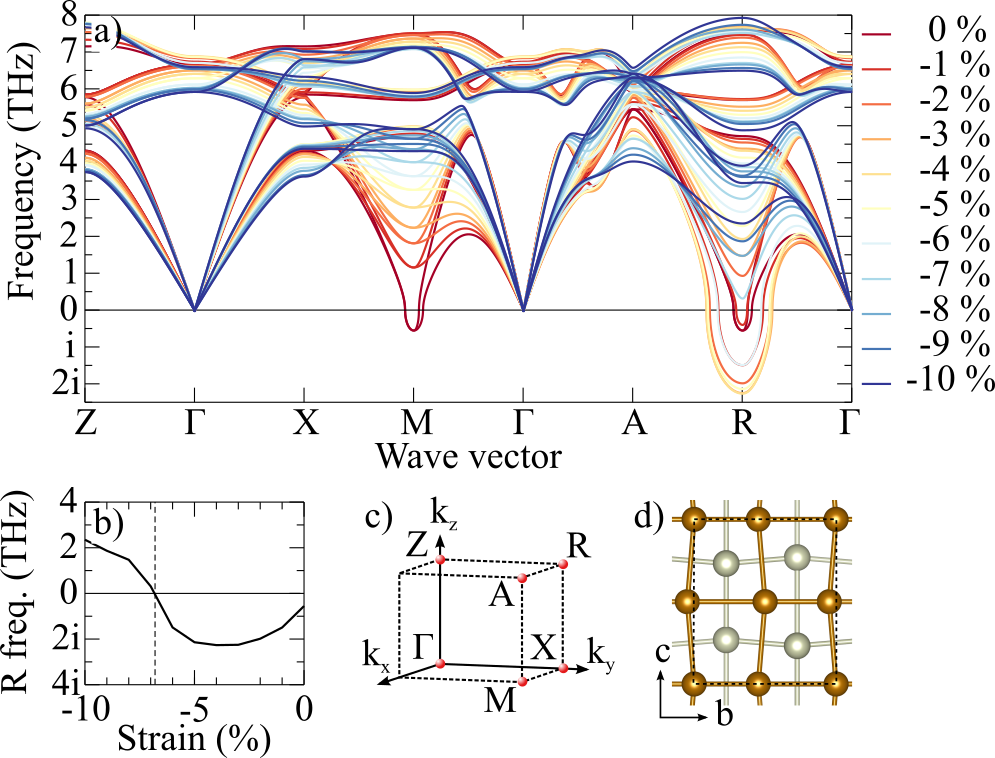}
\caption{\label{fig:phonon}a) Compressive strain dependence of the phonon dispersion in AFM FeRh computed at the GGA level. In b) we show the evolution of the frequency as a function of strain at the unstable R point. In c) and d) we show the high symmetry points in the tetragonal Brillouin zone and the resulting structural distortion respectively.}
\end{figure}

In addition to this elastic instability that manifests under compressive strain, our calculations reveal a lattice instability in the AFM phase corresponding to dimerisations of Fe and Rh atoms along perpendicular directions. In in Fig. \ref{fig:phonon}a) we show the evolution of the phonon spectrum, computed using GGA-PBE, as a function of strain. It can be seen that at 0\% strain a lattice instability exists at the degenerate M and R points that manifests as an imaginary phonon frequency. Compressive strain lifts the degeneracy of the M and R point and rapidly suppresses this instability at the M point (wave-vector in the compressively strained plane, see Fig. \ref{fig:phonon}c) while it is enhanced at the R point (wave-vector with a component along the elongated out-of-plane axis). In Fig. \ref{fig:phonon}b) we show the evolution of this R-point frequency with strain. The instability is initially enhanced up to compressive strains of 3 to 4\% and is then rapidly suppressed and vanishes for compressive strain slightly below 7\%.

The amount of strain where the instability is maximal as well as where it vanishes are in surprisingly good agreement with the position of the top of the energy barrier and the metastable minimum respectively of the energy curve in Fig. \ref{fig:strain}a). It is known that Bain path transitions can be linked to unstable phonon modes \cite{Grimvall:2012fw}, however in the present case, the instability appears only in a small portion of q-space around the R-point. Such zone-boundary instabilities cannot be invoked to explain a Bain path transition, which relies on unstable long wave-length phonons \cite{Grimvall:2012fw}. We have verified that no long wave-length instability exists along the R-$\Gamma$ line, which could mean that the strain match might be a coincidence. Nevertheless it is interesting to note that both the present R-point instability as well as an elastic $C'$ instability are represented by transverse [110] modes.

In Fig. \ref{fig:phonon}d) we show the structural distortion associated with this instability condensed at its most energy-lowering amplitude for the 5\% compressively strained structure. We see that it corresponds to a dimerisation of the Rh atoms along the film normal and a dimerisation of the Fe atoms in the strain plane. The distortion results in an energy lowering of 10.8 meV per FeRh formula unit at this strain value. In the unstrained cubic bcc structure the energy lowering is about two orders of magnitude smaller at only 0.125 meV per formula unit. This small energy scale suggests that the occurrence of this phenomenon should be limited to very low temperatures, consistent with this structure not yet having been observed in experiment. Compressive strain could however trigger the dimerisation at higher temperatures. In terms of electrical properties, the unstable phonon has the effect of slightly lowering the DOS at the Fermi energy, at 5\% compressive strain going from 0.6 states/eV/f.u. in the unmodulated structure to 0.5 states/eV/f.u. in the modulated structure.

\begin{figure}
\includegraphics[width=0.9\columnwidth]{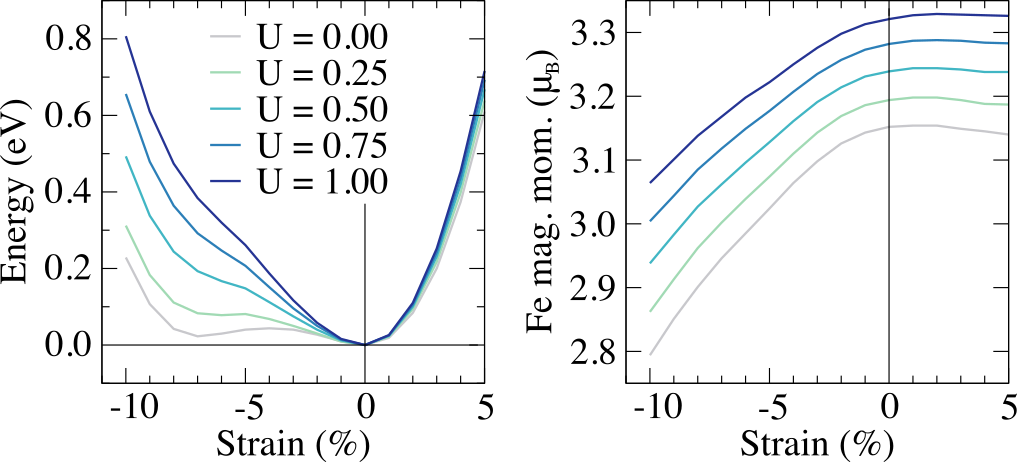}
\caption{\label{fig:U}a) GGA+U relative energy of the AFM phase with U values of 0, 1, 2 and 3 eV and b) the corresponding Fe magnetic moments.}
\end{figure}

Finally for this section, we present results of computer experiments to test the role of the size of the magnetic moments in determining the stability of the AFM phase with respect to epitaxial tetragonal strain. In order to check if a further increase of the moments would lead to a complete stabilisation of the AFM phase, we apply an on-site Hubbard U correction to the Fe d orbitals. We emphasize that we do not propose this as a good physical description of FeRh, but use it as a computational trick to tune the magnitude of the magnetic moments by modifying the charge localisation on the Fe atoms. In Fig. \ref{fig:U}a), we show the evolution of the energy as a function of strain for different values of U added to a GGA calculation. As shown above, without this artificial increase of the moments (U=0.00 eV), the secondary minimum of slightly higher energy exists for compressive strain. Upon increasing U, the secondary minimum disappears around U$\sim$0.25 eV, but remains visible as a kink in the energy curve. As is to be expected, increasing U increases the magnitude of the magnetic moments as shown in Fig. \ref{fig:U}b). Going from U=0 eV to 1 eV, increases the moments by 0.2 $\mu_B$ and more. This confirms indeed that increasing the magnetic moments will completely stabilise the CsCl structure with respect to tetragonal strain.

Our findings also confirm our assertion that the addition of a Hubbard U correction is not a physically appropriate description. As we mentioned in the introduction, experimental evidence supports the existence of an fcc phase, which can be reached by high energetic deformations of the material \cite{Lommel:1967cc,Miyajima:1992ta}. However, addition of a U > 0.25 to the GGA functional suppresses this state completely. This leads us to believe that semi-local functionals without any on-site correction are the best choice to describe the inter-metallic compound FeRh in a density functional theory calculation.

\subsection{Non-linear elasticity}

Finally we analyze the anomalous elastic behaviour of the AFM state under compressive epitaxial strain seen in Fig.~\ref{fig:strain}a in the framework of non-linear elasticity. In particular, we investigate how the linear elastic constant $C'=(C_{11}-C_{12})/2$ varies with strain, and which orders in the energy expansion we need to include in order to give a good description of the equilibrium AFM state and the surrounding saddle point region. Our main findings are first an unusually strong dependence of the linear elastic constants with strain, with $C'$ undergoing a strong softening under compression, and second that elastic constants up to fifth order are required to describe the elastic energy surface. These results confirm that the ground state of FeRh is proximal to an elastic instability. 

\begin{table}
\caption{AFM FeRh elastic constants obtained by fitting to DFT calculations of both orthorhombic and monoclinic distortions as a function of an initial isotropic distortion, and epitaxial-strain distortions. Slight deviations of the linear coefficients with respect to the values in Table \ref{tab:constants} stem from the different formalism and fitting procedure.} \label{TabEC}
\begin{ruledtabular}
\begin{tabular}{cc}
Elastic constants & AFM (GPa) \\
\hline
$C_{11}$, $C_{12}$, $C_{44}$ & 224.57, 184.60, 128.05 \\

$C_{111}$, $C_{112}$, $C_{123}$ & -2380.67, -561.34, -1461.22 \\
$C_{144}$, $C_{166}$, $C_{456}$ & 56424.12, -29201.20, - \\

$C_{1111}$, $C_{1112}$ & 27774.20, 646.69 \\
$C_{1122}$, $C_{1123}$ & 1206.44, 6456.56 \\
$C_{1144}$, $C_{1155}$ & 5875091.80, -2355521.45 \\
$C_{1255}$, $C_{1266}$ & 647062.69, -1860996.55 \\
$C_{1456}$, $C_{4444}$ & -, -71474.99 \\
$C_{4455}$ & - \\
$C_{11111}$, $C_{11112}$ & -444308.22, -22012.63 \\
$C_{11122}$, $C_{11123}$ & 39249.66, 60554.46 \\
$C_{11144}$, $C_{11155}$ & 252799931.07, -189779777.14 \\
$C_{11223}$, $C_{11244}$ & -79195.94, -10605443.05 \\
$C_{11255}$, $C_{11266}$ & 87944526.61, 61761708.62 \\
$C_{11456}$, $C_{11244}$ & -, -10605443.05 \\
$C_{12456}$, $C_{14444}$ & -, -501845412.25 \\
$C_{14455}$, $C_{15555}$ & -, 253346458.03 \\
$C_{15566}$, $C_{44456}$ & -, - \\
\end{tabular}
\end{ruledtabular}
\label{tab:nonlcst}
\end{table}

Using the GGA exchange-correlation functional that we established above to give the best description of FeRh, we performed two series of deformations to extract the elastic constants: First an isotropic volume distortion followed by either a volume-conserving monoclinic or orthorhombic distortion~\cite{Mehl:1994wu}. This allowed us to express the tension/compression asymmetry, which is normally fitted to a Murnaghan-Birch fit~\cite{Murnaghan:1944ch,Birch:1947da}, directly in terms of the higher order elastic constants: 
\begin{eqnarray}
E&=&\eta^2 \left(\frac{3 C_{11}}{2}+3 C_{12}\right)+\nonumber\\
& &\eta ^3 \left(\frac{C_{111}}{2}+3 C_{112}+C_{123}\right)+\nonumber\\
& &\eta ^4\left(\frac{C_{1111}}{8}+C_{1112}+\frac{3 C_{1122}}{4}+\frac{3 C_{1123}}{2}\right)+\label{EqnBM}\\
& &\eta ^5\left(\frac{C_{11111}}{40}+\frac{C_{11112}}{4}+\frac{C_{11122}}{2}+\frac{C_{11123}}{2}+\right.\nonumber\\
& &\qquad\left. \frac{3C_{11223}}{4}\right) \nonumber, 
\end{eqnarray}
where $\eta$ is the diagonal term of the Lagrangian strain matrix associated with isotropic expansion (for a review of non-linear elasticity theory see the Appendix). Second, we performed an isotropic in-plane distortion ($\varepsilon_{1}$) plus a perpendicular out-of-plane distortion ($\varepsilon_{2}$), corresponding to a biaxial strain state. We then fit the non-linear elastic energy expression, Eqn.~\ref{EqnEDcont}, to our DFT database of energies versus distortions using a simulated annealing algorithm~\cite{Corana:1987it}. The resulting optimal numerical values of the non-linear elastic constants are listed in Tab.~\ref{TabEC}. Note that the chosen distortions do not directly probe all elastic constants. For example $C_{456}$, $C_{1456}$, $C_{4455}$, $C_{11456}$, $C_{12456}$, $C_{14455}$, $C_{15566}$ and $C_{44456}$ do not contribute to the elastic energies and are therefore omitted from the fitting process, and others enter only in linear combinations (for example, see Eqns.~\ref{EqnBM} and \ref{EqnCp}). Thus caution should be taken in interpreting the numerical values of individual higher-order elastic constants. For conciseness we present results only for the AFM phase.

\begin{figure}
\includegraphics[width=0.9\columnwidth]{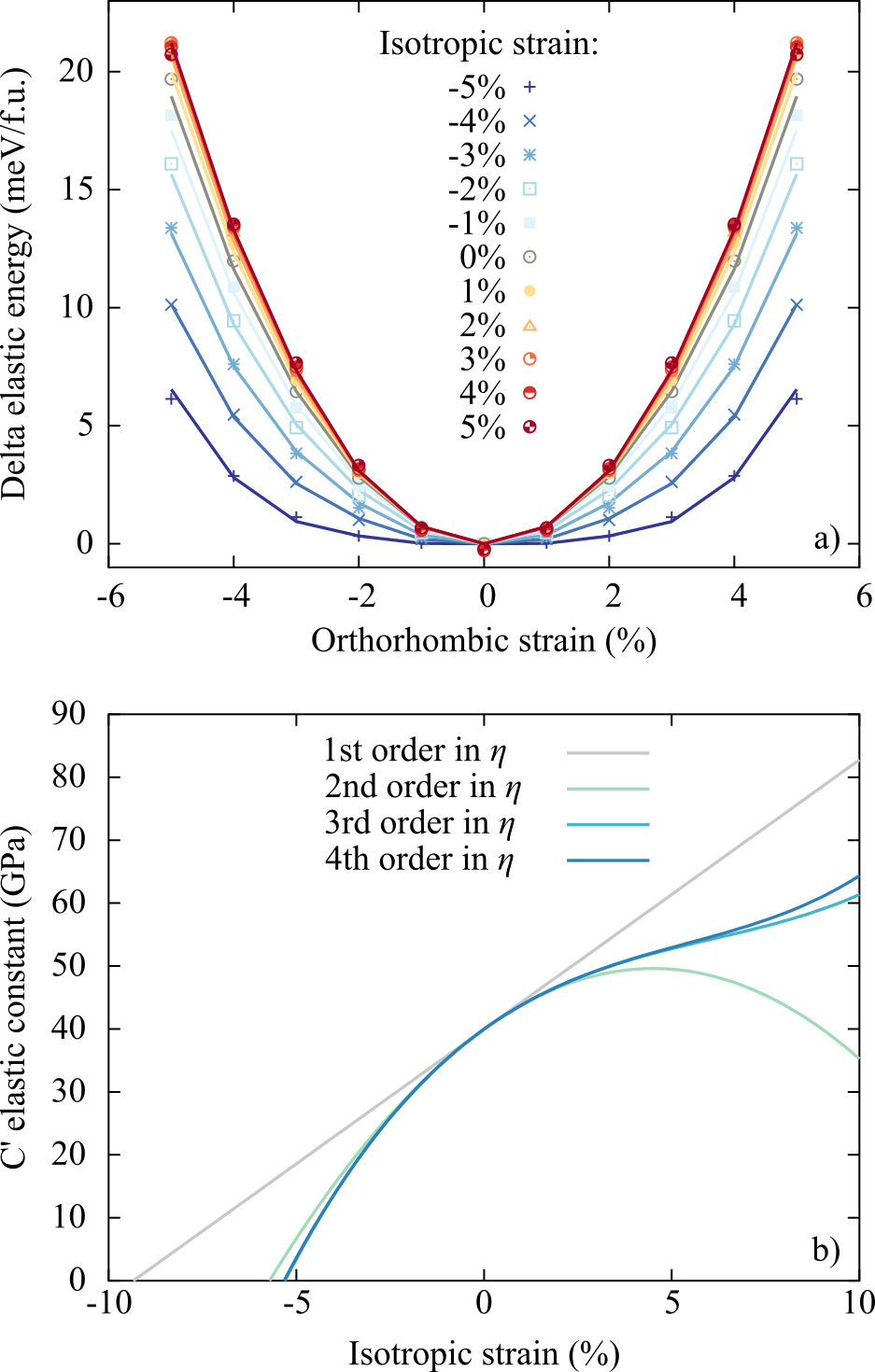}
\caption{a) Change in elastic energy (per formula unit) at a finite isotropic strain with respect to a volume conserving orthorhombic strain for the AFM FeRh system. For each color the points show the DFT data and the lines the eqn.~\ref{EqnEDcont} fit for a specific value of isotropic strain. b) Dependence of the AFM shear moduli $C'$ on isotropic strain according to eqn.~\ref{EqnCp}. Here different orders with respect to the isotropic strain are shown. Convergence under compression is achieved at fourth order.} \label{FigIsoMonoAFM}
\end{figure}

Fig.~\ref{FigIsoMonoAFM}a shows the change in elastic energy as a function of the volume-conserving orthorhombic distortion for different isotropic strains. Both the DFT data and the non-linear elasticity fit are shown, where the zero energy (at zero orthorhombic strain) is set to the AFM energy per formula unit of the given isotropic strain. We see that the $C'$ shear modulus changes as a function of isotropic strain (as does the $C_{44}$ shear modulus --- not shown). In particular there is a strong softening of the $C'$ modulus at negative isotropic strains (corresponding to lattice contraction). For the FM phase, no softening of the FM $C'$ modulus under isotropic compression was seen and for the NM phase, the $C'$ modulus was always negative indicating a broad regime of elastic instability as suggested by sec.~\ref{ss:ec}.

Insight into the origin of this behavior can be gained by writing down the isotropic strain dependence of the $C'$ elastic constant in terms of the higher order elastic constants. Up to quadratic order with respect to the isotropic strain, $\eta$, this is
\begin{eqnarray}
C'(\eta)&=&\frac{1}{2}\left(C_{11}-C_{12}\right)+\nonumber\\
& &\frac{1}{2}\left(6 C_{11}+C_{111}-C_{123}\right)\eta+ \label{EqnCp}\\
& &\frac{1}{2} \left(11C_{11}+4 C_{12}+\frac{11C_{111}}{2}+6 C_{112}-\right.\nonumber\\
& &\qquad\left.\frac{5 C_{123}}{2}+\frac{C_{1111}}{2}+C_{1112}-\frac{3C_{1123}}{2}\right) \eta^{2}+\dots \nonumber
\end{eqnarray}
A similar expression for the $C_{44}$ constant can also be obtained. Inspection of Eqn.~\ref{EqnCp} reveals that the higher order elastic constants strongly affect each coefficient of the $\eta$ expansion. Not shown are the $\eta^{3}$ and $\eta^{4}$ terms, which include respectively the fourth and fifth order order elastic constants. In Fig.~\ref{FigIsoMonoAFM}b we plot $C'$ including contributions up to fourth order in $\eta$ using the obtained numerical elastic constants for the AFM state for a range of negative and positive isotropic strains. We see again that, consistent with Fig. \ref{FigIsoMonoAFM}a, $C'$ softens on lattice contraction, eventually becoming negative at an isotropic strain of approximately -5\%. Thus under a sufficiently large isotropic compression the AFM state becomes unstable to an orthorhombic distortion. For the considered range of isotropic compressions, this trend is not seen in the FM phase.

Finally we note that the elastic constant behaviour of FeRh is remarkably similar to that of pure bcc Fe, which also exhibits a $C'$ softening to zero under compression, and has a negative $C'$ for the NM phase. For bcc Fe, this phenomenon is associated with the suppression of itinerant magnetism under compression, and it has been pointed out that this is a manifestation of the magneto-volume effect \cite{Herper:1999id, Guo:2000wc, Dudarev:2005ju}. As we pointed out earlier, (Fig.~\ref{fig:isostatic}) the energies of the AFM, FM and NM phases tend to the same values for sufficiently high compressions, and this is accompanied by a reduction in the size of the magnetic moments. For example, at an isotropic compression of 5\%, the individual Fe moment values of the AFM FeRh phase are lower than their equilibrium values by approximately 10\%.

\section{Conclusions}

In conclusion, we have shown that the GGA exchange-correlation functional provides a reasonable description of the known properties of FeRh, in contrast to the LDA which predicts incorrectly that a competing fcc-like phase with smaller lattice constant is the ground state, and the GGA+U method which incorrectly completely destabilizes this competing phase. At the GGA level the fcc-like state is metastable, consistent with reports of its existence under high-impact deformation, and we predict that it can be reached with compressive epitaxial strain.  Due to an enhanced density of states at the Fermi energy, this fcc-like state shows a decreased electrical resistivity compared to the bcc AFM phase. We rationalize the behavior using non-linear elasticity theory, and predict the existence of a lattice instability, which should manifest at low temperatures as a dimerisation of both Fe and Rh atoms along perpendicular directions. 

\section{Acknowledgements}

This work was financially supported by the ETH Z\"urich and by the ERC Advanced Grant program, No. 291151.
This work was supported by a grant from the Swiss National Supercomputing Centre (CSCS) under project ID s624.

\bibliography{references.bib}

\appendix

\section{Non-linear elasticity: definitions and fitting procedure} \label{app:nlelasticity}

Non-linear elasticity expands the elastic energy to supralinear order with respect to the deformation matrix, and thus defines higher than linear order elastic coefficients. Unlike linear elasticity, non-linear elasticity involves finite distortions of a material and therefore allows for the study of $C'$ and $C_{44}$ as a function of an arbitrary distortion. Here we outline non-linear elasticity theory to fifth order.

The higher order elastic constants depend on the precise definition of strain. Here the Lagrangian strain is used, which gives a measure of the finite displacement of a material point~\cite{Dunstan:2002kz,Johal:2006hm,Lopuszynski:2007el}. The Lagrangian strain is formally defined as
\begin{equation}
\boldsymbol{\eta}=\frac{1}{2}\left(\mathbf{J}\mathbf{J}^{T}-\mathbf{1}\right),
\end{equation}
where the Jacobian matrix, $\mathbf{J}$, is obtained from the linear strain tensor,
\begin{equation}
\mathbf{J}=\mathbf{1}+\boldsymbol{\varepsilon}.
\end{equation}
Thus the Lagrangian strain is
\begin{equation}
\boldsymbol{\eta}=\boldsymbol{\varepsilon}+\frac{1}{2}\boldsymbol{\varepsilon}^{2}. \label{EqnLS}
\end{equation}

In terms of Voigt notation, the most general elastic energy density is written as
\begin{equation}
E=E_{0}+\frac{1}{2!}\sum_{ij}C_{ij}\eta_{i}\eta_{j}+\frac{1}{3!}\sum_{ijk}C_{ijk}\eta_{i}\eta_{j}\eta_{k}+\dots.
\end{equation}
Here $\eta_{1}=\eta_{11}$, $\eta_{2}=\eta_{22}$, $\eta_{3}=\eta_{33}$, $\eta_{4}=2\eta_{23}$, $\eta_{5}=2\eta_{13}$ and $\eta_{6}=2\eta_{12}$. To fifth order, the elastic energy density becomes
\begin{equation}
E=\phi_{2}+\phi_{3}+\phi_{4}+\phi_{5}+\dots, \label{EqnEDcont}
\end{equation}
where for a cubic system the linear elastic contribution is given by
\begin{eqnarray}
\phi_{2}& =& \frac{C_{11}}{2} \left(\eta_{1}^{2} + \eta_{2}^{2} + \eta_{3}^{2}\right) +  \nonumber \\
& & \frac{C_{44}}{2} \left(\eta_{4}^{2} + \eta_{5}^{2} + \eta_{6}^{2}\right) + \nonumber \\
& & C_{12} \left(\eta_{1} \eta_{2} + \eta_{3} \eta_{2} + \eta_{1} \eta_{3}\right).\label{EqnED2}
\end{eqnarray}

The cubic higher order non-linear elastic energy terms are
\begin{widetext}
\begin{eqnarray}
\phi_{3}&=&\frac{C_{111}}{6} \left(\eta_{1}^{3} + \eta_{2}^{3} 
+ \eta_{3}^{3}\right) + \frac{C_{112}}{2} \left(\eta_{2} \eta_{1}^{2} + \eta_{3} \eta_{1}^{2} + \eta_{2}^{2} \eta_{1} + \eta_{3}^{2} \eta_{1} + 
\eta_{2} \eta_{3}^{2} + \eta_{2}^{2} \eta_{3}\right)  + C_{123} \eta_{1} \eta_{2} \eta_{3} + \nonumber \\
& &\frac{C_{144}}{2} \left(\eta_{1} \eta_{4}^{2} + \eta_{2} \eta_{5}^{2} + \eta_{3} \eta_{6}^{2}\right) + \frac{C_{166}}{2} \left(\eta_{2} \eta_{4}^{2} + \eta_{3} \eta_{4}^{2} + \eta_{1} \eta_{5}^{2} + \eta_{3} \eta_{5}^{2} + 
\eta_{1} \eta_{6}^{2} + \eta_{2} \eta_{6}^{2}\right) + 
C_{456} \eta_{4} \eta_{5} \eta_{6} \label{Eqn3O}
\end{eqnarray}

\begin{eqnarray}
\phi_{4} &= &\frac{C_{1111}}{24} \left(\eta_{1}^{4} + \eta_{2}^{4} + \eta_{3}^{4}\right) + \frac{C_{1112}}{6} \left(\eta_{1}^{3} \left(\eta_{2} + \eta_{3}\right) + \eta_{2}^{3} \left(\eta_{3} + \eta_{1}\right) + 
      \eta_{3}^{3} \left(\eta_{1} + \eta_{2}\right)\right) + \nonumber \\
& & \frac{C_{1122}}{4} \left(\eta_{1}^{2} \eta_{2}^{2} + \eta_{2}^{2} \eta_{3}^{2} + \eta_{3}^{2} \eta_{1}^{2}\right)+\frac{C_{1123}}{2} \left(\eta_{1}^{2} \eta_{2} \eta_{3} + \eta_{2}^{2} \eta_{3} \eta_{1} + \eta_{3}^{2} \eta_{1} \eta_{2}\right) + \nonumber \\ 
& &\frac{C_{1144}}{4} \left(\eta_{1}^{2} \eta_{4}^{2} + \eta_{2}^{2} \eta_{5}^{2} + \eta_{3}^{2} \eta_{6}^{2}\right) + \frac{C_{1155}}{4} \left(\eta_{1}^{2} \left(\eta_{5}^{2} + \eta_{6}^{2}\right) + \eta_{2}^{2} \left(\eta_{4}^{2} + \eta_{6}^{2}\right) + \eta_{3}^{2} \left(\eta_{4}^{2} + \eta_{5}^{2}\right)\right) + \nonumber \\
& & \frac{C_{1255}}{2} \left(\eta_{1} \eta_{2} \left(\eta_{4}^{2} + \eta_{5}^{2}\right) + \eta_{2} \eta_{3} \left(\eta_{5}^{2} + \eta_{6}^{2}\right) + \eta_{1} \eta_{3} \left(\eta_{4}^{2} + \eta_{6}^{2}\right)\right) +
\frac{C_{1266}}{2} \left(\eta_{1} \eta_{2} \eta_{6}^{2} + \eta_{2} \eta_{3} \eta_{4}^{2} + \eta_{1} \eta_{3} \eta_{5}^{2}\right) + \nonumber \\
& &
C_{1456} \eta_{4} \eta_{5} \eta_{6} \left(\eta_{1} + \eta_{2} + \eta_{3}\right) + 
\frac{C_{4444}}{24} \left(\eta_{4}^{4} + \eta_{5}^{4} + \eta_{6}^{4}\right) + 
\frac{C_{4455}}{4} \left(\eta_{4}^{2} \eta_{5}^{2} + \eta_{5}^{2} \eta_{6}^{2} + \eta_{6}^{2} \eta_{4}^{2}\right)\label{Eqn4O}
\end{eqnarray}
\begin{eqnarray}
\phi_{5}& =& \frac{C_{11111}}{120} \left(\eta_{1}^{5} + \eta_{2}^{5} + \eta_{3}^{5}\right) +  \frac{C_{11112}}{24} \left(\eta_{1}^{4} \left(\eta_{2} + \eta_{3}\right) + \eta_{2}^{4} \left(\eta_{1} + \eta_{3}\right) + 
      \eta_{3}^{4} \left(\eta_{1} + \eta_{2}\right)\right) + \nonumber \\
& &
   \frac{C_{11122}}{12} \left(\eta_{1}^{3} \left(\eta_{2}^{2} + \eta_{3}^{2}\right) + \eta_{2}^{3} \left(\eta_{1}^{2} + \eta_{3}^{2}\right) + 
      \eta_{3}^{3} \left(\eta_{1}^{2} + \eta_{2}^{2}\right)\right) + 
   \frac{C_{11123}}{6} \left(\eta_{1}^{3} \eta_{2} \eta_{3} + \eta_{2}^{3} \eta_{1} \eta_{3} + \eta_{3}^{3} \eta_{1} \eta_{2}\right) +\nonumber \\
& &
    \frac{C_{11144}}{12} \left(\eta_{1}^{3} \eta_{4}^{2} + \eta_{2}^{3} \eta_{5}^{2} + \eta_{3}^{3} \eta_{6}^{2}\right) + 
   \frac{C_{11155}}{12} \left(\eta_{1}^{3} \left(\eta_{5}^{2} + \eta_{6}^{2}\right) + \eta_{2}^{3} \left(\eta_{4}^{2} + \eta_{6}^{2}\right) + 
      \eta_{3}^{3} \left(\eta_{4}^{2} + \eta_{5}^{2}\right)\right) + \nonumber \\
& &
   \frac{C_{11223}}{4} \left(\eta_{1}^{2} \eta_{2}^{2} \eta_{3} + \eta_{1}^{2} \eta_{3}^{2} \eta_{2} + 
      \eta_{3}^{2} \eta_{2}^{2} \eta_{1}\right) + 
   \frac{C_{11244}}{4} \left(\eta_{1}^{2} \eta_{4}^{2} \left(\eta_{2} + \eta_{3}\right) + 
      \eta_{2}^{2} \eta_{5}^{2} \left(\eta_{1} + \eta_{3}\right) + \eta_{3}^{2} \eta_{6}^{2} \left(\eta_{1} + \eta_{2}\right)\right) +  \nonumber \\
& &
   \frac{C_{11255}}{4} \left(\eta_{1}^{2} \left(\eta_{2} \eta_{5}^{2} + \eta_{3} \eta_{6}^{2}\right) + 
      \eta_{2}^{2} \left(\eta_{1} \eta_{4}^{2} + \eta_{3} \eta_{6}^{2}\right) + 
      \eta_{3}^{2} \left(\eta_{1} \eta_{4}^{2} + \eta_{2} \eta_{5}^{2}\right)\right) + \nonumber \\
& &
   \frac{C_{11266}}{2} \left(\eta_{1} \eta_{2} \eta_{6}^{2} \left(\eta_{1} + \eta_{2}\right) + 
      \eta_{1} \eta_{3} \eta_{5}^{2} \left(\eta_{1} + \eta_{3}\right) + \eta_{2} \eta_{3} \eta_{4}^{2} \left(\eta_{2} + \eta_{3}\right)\right) + \nonumber \\
& &
   \frac{C_{11456}}{2} \eta_{4} \eta_{5} \eta_{6} \left(\eta_{1}^{2} + \eta_{2}^{2} + \eta_{3}^{2}\right) + 
   \frac{C_{12344}}{2} \eta_{1} \eta_{2} \eta_{3} \left(\eta_{4}^{2} + \eta_{5}^{2} + \eta_{6}^{2}\right) + \nonumber \\
& &
   C_{12456} \eta_{4} \eta_{5} \eta_{6} \left(\eta_{1} \eta_{2} + \eta_{1} \eta_{3} + \eta_{2} \eta_{3}\right) + 
   \frac{C_{14444}}{24} \left(\eta_{1} \eta_{4}^{4} + \eta_{2} \eta_{5}^{4} + \eta_{3} \eta_{6}^{4}\right) + \nonumber \\
& &
   \frac{C_{14455}}{4} \left(\eta_{4}^{2} \eta_{5}^{2} \left(\eta_{1} + \eta_{2}\right) + 
      \eta_{4}^{2} \eta_{6}^{2} \left(\eta_{1} + \eta_{3}\right) + \eta_{5}^{2} \eta_{6}^{2} \left(\eta_{2} + \eta_{3}\right)\right) + \nonumber \\
& &
   \frac{C_{15555}}{24} \left(\eta_{5}^{4} \left(\eta_{1} + \eta_{3}\right) + \eta_{4}^{4} \left(\eta_{2} + \eta_{3}\right) + 
      \eta_{6}^{4} \left(\eta_{1} + \eta_{2}\right)\right) + \nonumber \\
& &
   \frac{C_{15566}}{2} \left(\eta_{1} \eta_{5}^{2} \eta_{6}^{2} + \eta_{2} \eta_{4}^{2} \eta_{6}^{2} + 
      \eta_{3} \eta_{4}^{2} \eta_{5}^{2}\right) + 
   \frac{C_{44456}}{2} \eta_{4} \eta_{5} \eta_{6} \left(\eta_{4}^{2} + \eta_{5}^{2} + \eta_{6}^{2}\right)\label{Eqn5O}
\end{eqnarray}
\end{widetext}
Eqns.~\ref{Eqn3O} to \ref{Eqn5O} give the resulting expressions for the elastic energy density (Eqn.~\ref{EqnEDcont}) for the third, fourth and fifth order non-linear elastic contributions. Apart from a few typographical errors in the earlier papers, the expressions are similar to those found in Refs.~\cite{Ghate:1964ic,Chung:1974ga,Wang:2009ca}.

\end{document}